\documentclass[conference]{IEEEtran}
\IEEEoverridecommandlockouts

\usepackage{cite}
\usepackage{amsmath,amssymb,amsfonts}
\usepackage{algorithmic}
\usepackage{graphicx}
\usepackage{textcomp,color}
\usepackage{xcolor}
\usepackage{multirow}
\usepackage{threeparttable}
\usepackage{booktabs}
\usepackage{makecell}
\usepackage{hyperref}
\usepackage{enumitem}
\usepackage{amsmath}
\def\BibTeX{{\rm B\kern-.05em{\sc i\kern-.025em b}\kern-.08em
    T\kern-.1667em\lower.7ex\hbox{E}\kern-.125emX}}

\begin{document}

\title{Reducing the Gap Between Pretrained Speech Enhancement and Recognition Models \\ Using a Real Speech-Trained Bridging Module
}

\author{
\IEEEauthorblockN{
Zhongjian Cui$^1$, Chenrui Cui$^1$, Tianrui Wang$^1$, Mengnan He$^{2}$, Hao Shi$^3$, Meng Ge$^1$, \\ Caixia Gong$^2$, Longbiao Wang$^{1,4}$, Jianwu Dang$^5$\thanks{Mengnan He and Longbiao Wang are corresponding authors. This work was supported by the National Natural Science Foundation of China under Grant 62176182 and U23B2053.}
}
\vspace{0.1cm}
\IEEEauthorblockA{
\textit{$^1$Tianjin Key Laboratory of Cognitive Computing and Application,}
\textit{Tianjin University, Tianjin, China}\\
\textit{$^2$DiDi Chuxing, Beijing, China\quad}
\textit{$^3$Graduate School of Informatics, Kyoto University, Kyoto, Japan}\\
\textit{$^4$Huiyan Technology (Tianjin) Co., Ltd, Tianjin, China}\\
\textit{$^5$Shenzhen Institute of Advanced Technology, Chinese Academy of Sciences, Shenzhen, China}
}
}
\maketitle

\begin{abstract}
The information loss or distortion caused by single-channel speech enhancement (SE) harms the performance of automatic speech recognition (ASR). 
Observation addition (OA) is an effective post-processing method to improve ASR performance by balancing noisy and enhanced speech. 
Determining the OA coefficient is crucial. 
However, the currently supervised OA coefficient module, called the bridging module, only utilizes simulated noisy speech for training, which has a severe mismatch with real noisy speech. 
In this paper, we propose training strategies to train the bridging module with real noisy speech. 
First, DNSMOS is selected to evaluate the perceptual quality of real noisy speech with no need for the corresponding clean label to train the bridging module. 
Additional constraints during training are introduced to enhance the robustness of the bridging module further. 
Each utterance is evaluated by the ASR back-end using various OA coefficients to obtain the word error rates (WERs). 
The WERs are used to construct a multidimensional vector. 
This vector is introduced into the bridging module with multi-task learning and is used to determine the optimal OA coefficients.
The experimental results on the CHiME-4 dataset show that the proposed methods all had significant improvement compared with the simulated data trained bridging module, especially under real evaluation sets. 
\end{abstract}

\begin{IEEEkeywords}
Noise-robust automatic speech recognition, distortion, observation adding
\end{IEEEkeywords}

\section{Introduction}
Automatic speech recognition (ASR) aims to transcribe the text from speech \cite{10613503, kheddar2024automatic, koluguri2024investigating}. 
Its performance is greatly degraded with the presence of background noise \cite{li2014overview, rodrigues2019analyzing}. 
A common solution to improve the noise-robust ASR is to incorporate a speech enhancement (SE) front-end into the system. 
SE front-end can be classified into single-channel \cite{zhao2022frcrn, hu2020dccrn, wang2022hgcn, 10094718, wang2023harmonic, cao24_interspeech} and multi-channel \cite{heymann2016neural, boeddeker2018exploring, heymann2017beamnet, wang2021exploring} methods according to the channel number of the speech signals. 
Multi-channel front-ends often require more complex microphone arrays to pick up speech signals. 
Thus, compared with multi-channel microphone arrays, single-channel microphones are often more feasible to apply in practical scenarios. 
Nevertheless, single-channel SE front-ends often cause much more speech information distortion\cite{iwamoto2022bad, iwamoto2024does, ochiai2024rethinking}, which harms ASR performance.

The SE front-end and ASR back-end are often trained separately based on different learning targets. 
The inevitable mismatches between these two models lead to no improvement or degradation of ASR performance \cite{shi2024investigation, chen2018building}. 
Better integration of pre-trained SE front-end and ASR back-end is crucial to constructing a noise-robust ASR system. 
Joint training is an intuitive way to alleviate mismatches \cite{narayanan2014joint, wang2016joint, wu2017end, chen2018building, 10542371, chang2022end, shi2021spectrograms}. 
In this manner, the SE front-end typically acts as a feature extractor only for the corresponding ASR back-end, reducing the previously achieved speech noise reduction capability \cite{shi2021spectrograms}. 
Moreover, it is challenging to use joint training for noise-robust ASR when the front-end and back-end models cannot be trained or are unknown.

Existing studies \cite{iwamoto2022bad, iwamoto2024does, ochiai2024rethinking} find ASR back-ends are more sensitive to artifacts than noise. 
Observation addition (OA) is an effective post-processing to improve ASR performance \cite{sato2022learning}. 
It adds the original noisy speech to the enhanced speech with a certain coefficient. 
However, the OA coefficient should meet the diversity of the noise and signal-to-noise ratio (SNR) within different utterances \cite{lee2023lc4sv, chen2023noise}. 
The bridging module\cite{wang2024bridging}, which is according to the SNR or some other criteria, is used to select suitable OA coefficients for noise-robust ASR. 
The clean speech label for real noisy speech is unavailable, which makes the bridging module only trained with simulated noisy speech. 
However, there is a data distribution mismatch between the simulated and real speech. 
Thus, the bridging system trained with simulated data sometimes can not process real speech well.

\begin{figure*}[htb]
  \centering
  \includegraphics[width=0.8\linewidth]{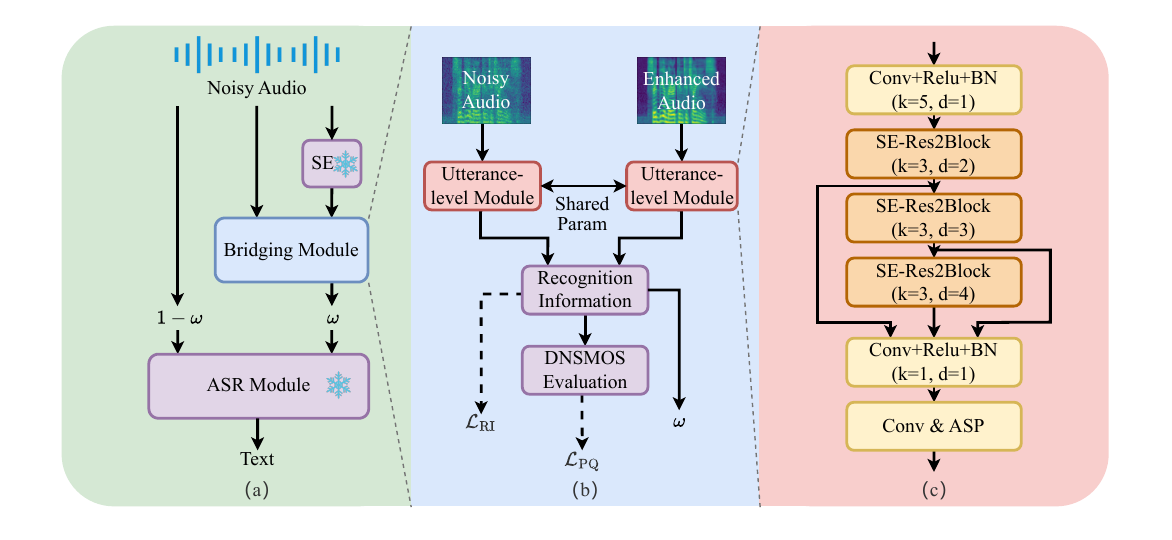}
  \vspace{-2.0em}
  \caption{The noise-robust ASR system based on bridging module (a) Overall framework (b) Bridging module structure: the recognition information module and perceptual quality module consist of a fully connected layer and a linear layer respectively (c) Utterance-level module structure}
  \label{fig1}
  \vspace{-1.5em}
\end{figure*}

In this paper, we propose training strategies to train the bridging module with the frozen SE and ASR model by using real noisy speech. 
Due to the success of training the bridging module according to noisy speech SNR on simulated data, speech evaluation metrics without clean labels required are considered. 
DNSMOS \cite{reddy2022dnsmos}, which measures speech quality without clean labels, is used as evaluation metrics. 
To match the OA coefficient and train the bridging module, the value estimated by DNSMOS is linearly mapped between 0 and 1. 
While effective for noise-robust ASR, DNSMOS has limitations, such as being overly smooth and less accurate for shorter clips. 
To further enhance the robustness of the bridging module, we introduce additional constraints during training.
It should be emphasized that ASR loss, i.e., joint training, is not used to train the bridging module.  
WER intuitively reflects whether the OA coefficient is appropriate for the utterance. 
Multiple OA coefficients are used to post-process and then compute WERs. 
For each utterance, a multidimensional vector reflects the WERs information according to different OA coefficients. 
The WERs information makes the bridging module learn a more suitable OA coefficient for the current ASR back-end with multi-task learning.

\section{Preparatory knowledge}

\subsection{Noise-Robust ASR with SE Front-end}
\label{AA}
Some studies \cite{li2022recent, wang2019overview} show that the SE front-end still benefits from the well-performing ASR performance. 
SE front-end aims to reduce the impact of noise on downstream tasks. 
It transforms the noisy speech signal $\boldsymbol{x}$ to enhanced speech signal $\boldsymbol{\widehat{y}}$, which can be expressed as follows: 
\vspace{-3pt}
\begin{equation}
\boldsymbol{\widehat{y}}=\mathrm{SE}(\boldsymbol{x}). 
\vspace{-3pt}
\end{equation}
Many network structures can be used to construct SE. 
Here, FRCRN \cite{zhao2022frcrn} and DCCRN \cite{hu2020dccrn} are chosen as the SE front-end. 
ASR back-end transcribes the enhanced speech signal $\boldsymbol{\widehat{y}}$ into word sequence $\boldsymbol{\widehat{W}}$, which can be expressed as follows: 

\vspace{-3pt}
\begin{equation}
\boldsymbol{\widehat{W}}=\mathrm{ASR}(\boldsymbol{\widehat{y}}). 
\vspace{-3pt}
\end{equation}
Whisper \cite{radford2022robustspeechrecognitionlargescale} is chosen as the ASR back-end. 

\subsection{Observation Addition}
The artifacts caused by single-channel SE affect ASR performance\cite{iwamoto2024does, ochiai2024rethinking, iwamoto2022bad}. 
OA is a simple and feasible way to alleviate signal artifacts. 
It adds the original noisy speech signal to the enhanced speech signal, which can be expressed as follows: 

\vspace{-3pt}
\begin{equation}
\boldsymbol{\tilde y}=\omega \boldsymbol{x}+(1-\omega)\boldsymbol{\widehat{y}},
\vspace{-3pt}
\end{equation}
where $\omega$ is used to balance the noisy and the enhanced speech signals, which is a pre-set value between 0 and 1.

\subsection{Bridging Module}
The existing bridging module \cite{wang2024bridging} is based on SNR. 
It first predicts the SNR-level of the speech through cosine similarity and then maps the OA coefficient through the linear layer. 
Finally, the bridging module is trained using the MSE loss between clean speech signals and the speech balanced by the noisy speech and enhanced speech with the OA coefficient.
The lower part of Fig.~\ref{fig1} (a) shows the OA with the bridging module. 
The bridging module is trained separately from the SE front-end and ASR back-end. 
The bridging module aims to establish a connection between the noisy and the enhanced speech signals to dynamically obtain a more matching $\omega$ for different speeches, rather than a fixed pre-set value: 

\vspace{-3pt}
\begin{equation}
\widehat{\omega}=\mathrm{Bridging}(\boldsymbol{x},\boldsymbol{\widehat{y}}),
\vspace{-3pt}
\label{eq:bridging}
\end{equation}
where $\widehat{\omega}$ is predicted by the bridging module.

\section{Real Speech-training for Bridging Module}
Fig. 1 illustrates our proposed method: a sentence-level bridging module, supervised by recognition information and perceptual quality. This module predicts OA coefficients from noisy and enhanced speech to generate inputs optimized for ASR.
\subsection{Training Strategies}
\subsubsection{Training with Perceptual Quality}

The impact of noise on speech is most directly reflected in perception. Therefore, we introduce non-intrusive perceptual metrics to guide the bridging module for predicting OA on real data. Specifically, we use the normalized $sig$ and $bak$ metrics from DNSMOS as the targets for the bridging module's prediction, as follows:

\vspace{-3pt}
\begin{equation}
\mathcal{L}_{\mathrm{PQ}}=MSE(\hat{\omega},(norm(sig)+norm(bak))/2),
\vspace{-3pt}
\end{equation}
where $norm$ represents the normalization of metric ranges from 1 to 5, as $norm(\alpha)=\frac{\alpha-1}{4}$. The $sig$ metric reflects the quality of the speech, while the $bak$ metric reflects the quality of the background noise. $\hat{\omega}$ is the predicted result from the bridging module, and $MSE$ represents the mean squared error loss function.

\subsubsection{Training with Recognition Information} 

In addition to perceptual quality, factors such as artifacts, noise residue, speech length, and disruptions in speech structure significantly affect the performance of speech recognition models\cite{shanthamallappa2024robust, ochiai2024rethinking, iwamoto2022bad, iwamoto2024does}, leading to the impact of word error rate (WER). Thus, we use WER directly as the second supervision for the bridging module. Specifically, we preset $(1/k+1)$ fixed OA coefficients with a step size of $k\in (0,0.1]$ from 0 to 1, applying these OA coefficients to the SE and ASR systems to process the training set. For each speech sample, we calculate $(1/k+1)$ WER values. The precomputed OA-WER pairs are then used to guide the training of the bridging module, as follows:
\vspace{-3pt}
\begin{equation}
\mathcal{L}_{RI}=-\log\sigma(cs(\sigma(logits),\sigma(wers))),
\vspace{-3pt}
\end{equation}
where $logits$ represent an intermediate representation of the bridging module, $\sigma$ is the sigmoid function, and $cs$ denotes cosine similarity. $wers = [wer_0, wer_1, \dots, wer_{(1/k)}]$ are the WER values for each speech sample, arranged in descending order based on OA. This allows $\mathcal{L}_{RI}$ to guide the bridging module in predicting the sigmoid-compressed OA-WER distribution, enabling the module to anticipate how different OA values affect speech recognition performance within the current SE and ASR systems.

\subsubsection{Combined Training}
Training the bridging module with both perceptual quality and recognition information is described as: 

\vspace{-3pt}
\begin{equation}
\widehat{\boldsymbol{\Lambda}}=\underset{{\boldsymbol{\Lambda}=\{\boldsymbol{\theta}_{{\mathrm{BM}}}\}}}{\operatorname*{\operatorname*{\operatorname*{argmin}}}}({\mathcal{L}_{{\mathrm{PQ}}}+\mathcal{L}_{{\mathrm{RI}}}})/2,
\vspace{-3pt}
\end{equation}

\noindent{where $\boldsymbol{\Lambda}$ represents the model parameters to be optimized. $\boldsymbol{\theta}_{{\mathrm{BM}}}$ represents the parameters of the bridging module. $\mathcal{L}_{{\mathrm{PQ}}}$ and $\mathcal{L}_{{\mathrm{RI}}}$ represent formula (5) and formula (6)}

\subsection{Model Structure}

Most ASR models use Fbank features for recognition. To adapt to these models, we extracted Fbank features as inputs for the bridging module. 
Considering that certain speech properties can be extracted from different sets of frames, we employ a channel-time attention method that leverages dependencies in the time dimension to comprehensively extract speech features. 
Specifically, we adopt the architecture of ResNet and use convolution and Attentive Stat Pooling (ASP) \cite{okabe18_interspeech}. 
According to \cite{iwamoto2022bad, iwamoto2024does, ochiai2024rethinking}, the OA coefficient primarily depends on two factors: noise in noisy speech and artifacts in enhanced speech. 
Consequently, we use features from both noisy and enhanced speech as inputs to the module, and achieve an OA coefficient that balances the ratio between noisy and enhanced speech through the perceptual quality and recognition information. The specific module structure is illustrated in Fig.~\ref{fig1} (b) and (c).

\section{Experimental}
\subsection{Dataset}
The experiments are conducted using the CHiME-4 dataset \cite{vincent20164th}. 
The training set includes 1,600 real and 7,138 simulated noisy speech samples, while the validation set comprises 1,640 real and their simulated noisy counterparts. 
The evaluation set consists of 1,320 real and 1,320 simulated noisy samples. 
Only Channel~5 is used for training, validation, and testing.

\begin{table*}[]
\centering
\caption{The WER of different pretrained SE models and different whispers on chime4.}
\vspace{-1.0em}
\renewcommand{\arraystretch}{1.1}
\begin{tabular}{c|c|ccc|ccc|ccc}
\bottomrule
\multirow{3}{*}{SE model} & \multirow{2}{*}{ASR (Whisper)}            & \multicolumn{3}{c|}{et\_simu} & \multicolumn{3}{c|}{et\_real} & \multicolumn{3}{c}{overall}  \\[0.1ex]
\cline{3-11}
                          &                                      & base    & small   & large-v3 & base    & small   & large-v3 & base    & small   & large-v3 \\[0.1ex]
\cline{2-11}
                          & Only ASR                             & 12.40\% & 7.51\%  & 4.35\%   & 18.24\% & 10.05\% & 4.99\%   & 15.32\% & 8.78\%  & 4.66\%   \\[0.1ex]
\cline{1-11}
\multirow{6}{*}{FRCRN}    & SE-ASR                               & 12.70\% & 11.29\% & 5.28\%   & 17.69\% & 14.02\% & 6.89\%   & 15.20\% & 12.66\% & 6.19\%   \\[0.1ex]
                          & SE-OA-ASR         & \pmb{10.41\%} & 9.59\%  & 4.24\%   & 17.80\% & 10.16\% & 4.78\%   & 14.10\% & 9.88\%  & 4.51\%   \\
                          & SE-BM-ASR(SNR-level)  & 12.18\% & 10.08\% & 4.24\%   & 17.37\% & 9.76\%  & 4.82\%   & 14.78\% & 9.92\%  & 4.53\%   \\
                          & SE-BM(ours)-ASR(SNR-level) & 12.59\% &
                          11.27\% & 5.45\% & 19.01\% & 13.93\% &
                          6.90\% & 15.80\% & 12.60\% & 6.18\% \\
                          & SE-BM-ASR($\mathcal{L}_{RI}$)             & 10.43\% & 9.60\%  & \underline{\pmb{4.07\%}}   & \underline{16.64\%} & 10.02\% & \underline{4.61\%}   & \underline{13.53\%} & 9.81\%  & \underline{4.30\%}   \\
                          & SE-BM-ASR($\mathcal{L}_{PQ}$)          & 11.03\% & \pmb{8.42\%}  & \underline{\pmb{4.07\%}}   & \underline{15.57\%} & \underline{8.79\%}  & \underline{4.52\%}   & \underline{13.30\%} & \underline{\pmb{8.61\%}}  & \underline{4.27\%}   \\
                          & SE-BM-ASR($\mathcal{L}_{RI}$ + $\mathcal{L}_{PQ}$) & 10.95\% & 9.55\%  & \underline{4.14\%}   & \underline{\pmb{15.52\%}} & \underline{\pmb{8.67\%}}  & \underline{\pmb{4.36\%}}   & \underline{\pmb{13.23\%}} & 9.11\%  & \underline{\pmb{4.25\%}}   \\
\cline{1-11}
\multirow{6}{*}{DCCRN}    & SE-ASR                               & 16.63\% & 12.07\% & 5.94\%   & 27.17\% & 17.83\% & 8.79\%   & 21.90\% & 14.95\% & 7.37\%   \\[0.1ex]
                          & SE-OA-ASR    & 12.24\% & 8.61\%  & 4.54\%   & 19.14\% & 9.80\%  & 5.25\%   & 15.69\% & 9.21\%  & 4.89\%   \\
                          & SE-BM-ASR(SNR-level)  & 15.93\% & 11.02\% & 5.87\%   & 26.46\% & 16.12\% & 7.87\%   & 21.20\% & 13.57\% & 6.87\%   \\
                          & SE-BM(ours)-ASR(SNR-level) & 16.62\% & 
                          12.06\% & 6.11\% & 29.04\% & 17.34\% & 8.74\% 
                          & 22.83\% & 14.70\% & 7.46\%\\
                          & SE-BM-ASR($\mathcal{L}_{RI}$)             & \underline{11.74\%} & 8.69\%  & \underline{4.30\%}   & \underline{17.76\%} & \underline{9.48\%}  & \underline{4.89\%}   & \underline{14.62\%} & 9.08\%  & \underline{4.59\%}   \\
                          & SE-BM-ASR($\mathcal{L}_{PQ}$)          & \underline{\pmb{11.42\%}} & 8.51\%  & \underline{4.21\%}   & \underline{17.19\%} & \underline{\pmb{9.38\%}}  & \underline{\pmb{4.76\%}}   & \underline{14.30\%} & 8.95\%  & \underline{4.49\%}   \\
                          & SE-BM-ASR($\mathcal{L}_{RI}$ + $\mathcal{L}_{PQ}$) & \underline{11.49\%} & \underline{\pmb{7.47\%}}  & \underline{\pmb{4.18\%}}   & \underline{\pmb{16.28\%}} & 9.81\%  & \underline{\pmb{4.76\%}}   & \underline{\pmb{13.98\%}} & \underline{\pmb{8.64\%}}  & \underline{\pmb{4.47\%}}   \\  
\bottomrule
\multicolumn{11}{c}{\textbf{Bold} denotes the best performance for each fixed SE and ASR model, and \underline{underline} indicates that the results outperform the baselines.} \\
\end{tabular}
\label{main_table}
\vspace{-1.5em}
\end{table*}

\subsection{Models}
This paper aims to use the bridging module to bridging the well-trained SE and ASR models. 
Thus, the parameters of SE and ASR models are frozen. 

\begin{enumerate}[topsep=0pt, fullwidth, itemindent=0em]
\setlength{\itemsep}{0pt}
\setlength{\parskip}{0pt}
\setlength{\parsep}{0pt}
\item[-] SE model: FRCRN and DCCRN are used. 

\item[-] ASR model: Whisper base, smalll, and large-v3 are used. 

\item[-] Bridging module: 
The input speech is converted into an 80-dimensional Fbank using a 25-ms window and a 10-ms frame shift. 
For data augmentation, we applied SpecAugment \cite{park2019specaugment}, which randomly masks 0 to 5 frames in the time domain and 0 to 4 channels in the frequency domain. 
The number of channels in the convolutional frame layer is set to 256 or 384. 
The bottleneck dimension in the SE-Block and attention module is set to 256. 
The scale dimension $s$ in Res2Block \cite{gao2019res2net} is set to 8. 
The number of nodes in the fully connected layer is set to 384. The fixed OA coefficient step size $k$ is set to 0.1.

\item[-] Training configuration: 
The learning rate is 0.0005. 
Additionally, we employ a warmup supervision, with the maximum number of training epochs set to 45. 
To prevent the issue of gradient explosion, the maximum norm for gradient clipping is set to 10.0.

\item[-] Baselines: 
Only ASR \cite{radford2022robustspeechrecognitionlargescale}, SE-ASR \cite{zhao2022frcrn}, SE-OA-ASR \cite{iwamoto2024does}, and SE-BM-ASR (SNR-level) \cite{wang2024bridging} are baselines. 
\end{enumerate}



\begin{table}[]
\vspace{-0.5em}
\caption{The WER of frcrn and whisper small on chime4 with different OA coefficient.}
\centering
\vspace{-1.0em}
\renewcommand{\arraystretch}{1.1}
\begin{tabular}{cc|cc|cc|cc}
\bottomrule
OA  & WER     & OA  & WER    & OA  & WER    & OA  & WER    \\
\cline{1-8}
0   & 12.07\% & 0.3 & 8.79\% & 0.6 & 8.53\% & 0.9 & 8.61\% \\
0.1 & 11.16\% & 0.4 & 8.80\% & 0.7 & 8.53\% & 1.0 & 7.51\% \\
0.2 & 9.14\%  & 0.5 & 8.54\% & 0.8 & 8.56\% &     &       \\
\toprule
\end{tabular}
\label{table_2}
\vspace{-2.5em}
\end{table}

\subsection{Experiments Result}
Table~\ref{main_table} presents the performance indicators regarding WER. 
\subsubsection{Importance of OA}
Most ``SE-ASR" combinations perform worse than ``Only ASR", indicating that the distortion introduced by SE reduces the performance of ASR.
In the ``SE-OA-ASR" combinations, we use the optimal OA coefficient from the validation set for the evaluation set. 
However, this coefficient does not consistently outperform the ``Only ASR" model, such as ``FRCRN-OA-Whisper small", suggesting that relying on the validation set's fixed OA coefficient is unreliable.  
Moreover, as shown in Table~\ref{table_2}, for the simulated evaluation data with DCCRN as front-end, all OA results are worse than using the ASR model alone, indicating that a fixed OA coefficient is not appropriate.

\subsubsection{Comparison of Different Supervision for BM}
We compare three of our proposed strategies (perceptual quality, recognition information, and combined training) with SNR-level supervision based on our BM module. 
Even though SNR-level sometimes have a positive effect, the effect is not significant. 
Perceptual quality and recognition information in our training strategies effectively enhance the overall performance. 
The perceptual quality captures optimal OA coefficients compared to recognition information. 
Our experiment demonstrates that the two training strategies, perceptual quality and recognition information, have distinct focuses, resulting in different outcomes when applied individually. 
However, the combined strategy effectively compensates for the limitations of these two strategies and further improves the performance, especially on real data. 

\begin{figure}[htb]
  \vspace{-0.5em}
  \centering
  \includegraphics[width=1.0\linewidth]{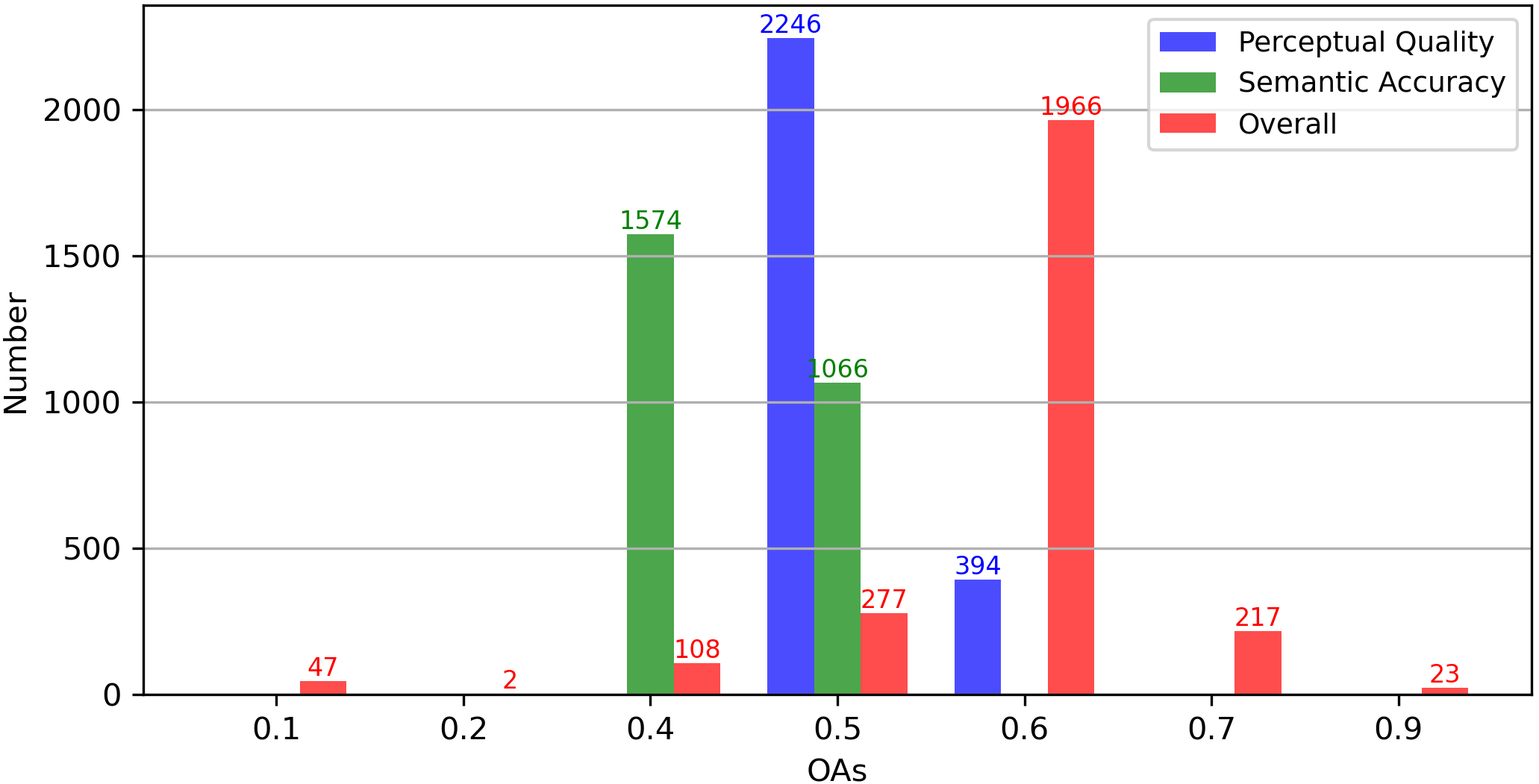}
  \vspace{-22pt}
  \caption{The number of utterances corresponding to different OA coefficients in the three training strategies}
  \label{fig3}
  \vspace{-1.6em}
\end{figure}

We analyze the OA distribution with the lowest WER for this model combination and found that it mainly falls between 0.4 and 0.7. 
As shown in Fig.~\ref{fig3}, we present the OA coefficient statistics derived from these three training strategies, which used the FRCRN as the SE model and the Whisper large-v3 as the ASR model. 
We find that perceptual quality focuses the OA distribution between 0.5 and 0.6, while recognition information concentrates it between 0.4 and 0.5. The combined strategy effectively integrates the strengths of these two strategies, enabling a more comprehensive selection of the optimal OA coefficient and achieving better performance.


\subsubsection{Comparison of Different ASR within Our Strategy}
The performance disparity of using only ASR between the base and the large-v3 is 13.25\%. 
After adding the DCCRN as the SE model, the performance disparity increases to 18.38\%.
However, after introducing the proposed bridging module and combined strategy, the performance disparity drops to 11.16\%. 
It demonstrates that our proposed bridging module and training strategies have greater potential for improvement, especially when training relatively weaker ASR models.

\subsubsection{Comparison of Different SE within Our Strategy}
The poor-performance SE model degrades the ASR's performance more. 
However, our proposed bridging module and combined strategy effectively narrowed this performance disparity. 
For instance, using the Whisper base model as an example, the performance disparity between FRCRN and DCCRN is 6.7\% before applying the proposed bridging module and combined strategy. 
After employing the proposed bridging module and combined strategy, this performance disparity 
between FRCRN and DCCRN 
is reduced to only 0.75\%. 
It demonstrates that our training strategies have more potential for improvement compared to weak-performance SE models.


\section{Conclusions}

In this paper, we propose three training strategies for the bridging module, focusing on perceptual quality, recognition information, and a combination of both, using real noisy speech without simulation speech. Systems trained with these strategies significantly enhance the noise robustness of ASR with frozen SE and ASR models. Our results demonstrate that the systems trained using our strategies significantly outperform reference methods. Moreover, our strategies substantially reduce the performance disparity between models with varying effectiveness levels, with the improvement being particularly prominent in the SE model. However, further optimization is needed for the use of WER in recognition information. We will focus on refining the recognition information strategy and better aligning it with the needs of the ASR models to improve the performance of the bridging module in future work. 


\clearpage

\bibliographystyle{IEEEtran}
\bibliography{reference}

\end{document}